# Superconducting single X-ray photon detector based on $W_{0.8}Si_{0.2}$

Xiaofu Zhang,[a] Qiang Wang and Andreas Schilling

*Physik-Institut, Universität Zürich, Winterthurerstr. 190, 8057 Zürich, Switzerland*

We fabricated a superconducting single X-ray photon detector based on $W_{0.8}Si_{0.2}$, and we characterized its basic detection performance for keV-photons at different temperatures. The detector has a critical temperature of 4.97 K, and it is able to be operated up to 4.8 K, just below the critical temperature. The detector starts to react to X-ray photons at relatively low bias currents, less than 1% of $I_c$ at $T$ = 1.8 K, and it shows a saturated count rate dependence on bias current at all temperatures, indicating that the optimum internal quantum efficiency can always be reached. Dark counts are negligible up to the highest investigated bias currents (99% of $I_c$) and operating temperature (4.8 K). The latching effect affects the detector performance at all temperatures due to the fast recovery of the bias current; however, further modifications of the device geometry are expected to reduce the tendency for latching.

Ultrafast single photon sensitive X-ray detectors have a potential for important applications in many areas. Using photon-counting detectors (PCDs), the image quality of the medical X-ray computed tomography (CT) with low X-ray dose can be significantly improved.[1, 2] Such applications can even be extended to single molecular, virus, or cell CT and X-ray imaging.[3] The currently used energy integrating detectors (EIDs) in CT scanners and X-ray systems, however, have certain limits with regard to this technology. The EIDs measure the energy integrated signals of X-ray photons,[1, 4] and they are affected by the electronic noise and Swank noise.[5] As a result, the weight of low energy photons is decreased, which in turn leads to an increase of noise and a decrease of contrast.[1] Moreover, the performance of PCDs based on semiconductor technology is not impeccable as the respective count rate is limited.[1, 4, 6-7] These detectors have a typical dead time of several hundred nanoseconds, which limits the maximum count rate per pixel to a few megahertz. Meanwhile, the pixel size of these detectors is of the order of several hundred micrometers, which results in a maximum counts-per-second-per-square-millimeters (CPSPSM) of $10^6$ cps.[1] The required count rate for a clinical X-ray CT scanner, however, may be as high as $10^9$ cps.[6] Apart from the medical applications, ultrafast single X-ray photon detector can also be used in synchrotron X-ray sources, free-electron lasers, and astronomy. Synchrotron radiation, for example, has provided the possibility to perform X-ray experiments at very short time scales, and the use of time resolving detectors is therefore essential.[8-10] The currently used

[a]zhang@physik.uzh.ch

Fast-Readout Low-Noise (FReLoN) detectors, however, have relatively large time jitter (more than ten nanoseconds).[10, 11] As a result, the highest possible time resolution in these experiments is limited.

We recently fabricated thick film superconducting nanowire single X-ray photon detectors (X-SNSPDs) based on Nb and TaN.[12, 13] The area of these detectors is around several thousand square micrometers, and the dead time is as low as several nanoseconds. Hence the resulting maximum CPSPSM can be as high as $10^{11}$ cps, nearly two orders of magnitude higher than the demands from clinical X-ray CT scanner, which would make this type of detectors excellently suitable for a CT application. Furthermore, compared with semiconductor detectors, the time jitter of typical optical SNSPDs can be as low as several ten picoseconds,[14, 15] and thus we expect that the jitter of X-SNSPDs is within at least 1 ns, which might significantly improve the time resolution in a corresponding X-ray experiment. The performance of these detectors, however, is still hampered by the low X-ray photon absorption in the used superconducting materials. Moreover, in order to have a good detection performance, these detectors need to be operated at relatively low temperature ($T$ = 1.85 K), and only narrow nanowire detectors have shown a saturated detection characteristics on varying bias current, which is necessary to reach optimum internal quantum efficiency. For wider superconducting strips with the same materials, no saturation was observed throughout the experiments.[12, 13] In earlier experiments, Perez de Lara et al.[16] used a 5 nm thick NbN based optical SNSPD for the 6 keV X-ray photon detection, but the corresponding count rates did not saturate at all, and the dark count rates were even comparable with the photon count rates, making this type of detector unsuitable for applications.

In this letter, we introduce the amorphous WSi materials ($W_{0.8}Si_{0.2}$) into the single X-ray photon detector fabrication, which have already shown great potential for single optical photon detector technology.[17-19] The X-ray photon absorption is expected to be enhanced by the heavy tungsten element, which in turn will improve the system detection efficiency.[20] Moreover, a larger hotspot is expected to be created in $W_{0.8}Si_{0.2}$ upon photon absorption due to the smaller superconducting energy gap in WSi, and therefore a saturated count rate as a function of the bias current might also be realized in wide-strip WSi detectors. Finally, as a consequence of the amorphous nature of WSi, a more homogeneous superconducting strip can be fabricated. Therefore, such a detector is likely to be able to function at relatively high temperatures, since uniform strips are more robust against the thermal fluctuations.[18]

The superconducting amorphous $W_{0.8}Si_{0.2}$ films were grown on silicon substrates by means of dc magnetron sputtering of a pure W target and rf magnetron sputtering of a pure Si target in argon atmosphere, at a total pressure of 3 mTorr.[21] The narrow strips were patterned by electron beam lithography with hydrogen silsesquioxane (HSQ) resist and etched by reactive ion etching in $SF_6$. The device has a strip width of $w$ = 920 nm with a filling factor of 66%, which covers a sensitive area of

$41.6 \times 28$ µm$^2$, as it is shown as a inset in Fig. 1(a). Figure 1(b) shows the superconducting transition of the device, and we fitted the temperature dependence of resistance with the three-dimension Aslamazov-Larkin fluctuation formula.[22] The thus obtained critical temperature is $T_c = 4.97$ K, which is identical to the as-grown films. Therefore the fabrication process does not have any detectable influence on $T_c$. The experimental critical current $I_c$, which was obtained from the I-V characteristic curve without X-ray illumination, is plotted as a function of temperature in Fig 1(c), and the data can be well fitted by the Ginzburg-Landau (GL) mean field expression.[23] The corresponding value of the critical temperature from this fit was $T_c = 5.02$ K, consistent with the result from the resistance measurements.

The detection performance was measured in a *Janis* He-3 bath cryostat, and the X-ray photons coming from a W-target X-ray tube (see in Fig. 1(a)) were introduced onto the detector through a 100 µm thick Kapton film on the cryostat window through free space coupling. The corresponding characteristic X-ray energies for the X-ray source at acceleration voltage $V_A = 30$ kV and 50 kV are $L_\alpha = 8.4$ keV, $L_{\beta 1} = 9.67$ keV, $L_{\beta 2} = 9.96$ keV, and $L_\gamma = 11.29$ keV. A 1 µm thick aluminium filter at $T = 77$ K is placed on the optical path to shield the infrared radiation from the X-ray source. A schematic diagram of our measurement system is shown in Fig. 1(a). The detector was biased under a constant current through a series of low-pass filters at room temperature and a bias-tee in the cryostat. The detection signal was firstly transmitted to a cryogenic amplifier at the 4 K-stage and further amplified by two amplifiers at room temperature with a total gain of 60 dB, and finally counted by a 3.5 GHz *Tektronix* digital oscilloscope. Due to the broader wire (see Table I), the signal pulse has a rise time around 400 ps, which is slightly larger than that of the relatively narrower Nb and TaN X-SNSPD. By changing the X-ray source current $I_x$ and acceleration voltage, the photon flux could be adjusted. Figure 2 shows the count-rate dependences (at $T = 2$ K) on the source current $I_x$, which is a measure of the X-ray photon flux. In order to guide the eye, we embed an artificial, dash line with a slope of 1, which is parallel to the measured data. The slope of the count rate as a function of the X-ray source current indicates that the detector works in the single-photon detection regime.[12, 13]

Figure 3 shows the count rates as functions of the bias current, normalized by the critical current at each temperature, for $V_A = 30$ kV and 49.9 kV. The X-ray source current $I_x$ has been fixed at 1 mA throughout the experiments. Our detector starts to react to the X-ray photons under a bias current as low as $I_b = 5$ µA in the low temperature range, i.e., well below 1% of $I_c$ at $T = 1.8$ K. In a typical optical SNSPD, detection events are collected at relatively high bias current, ~ 20% of $I_c$, and saturated count rate (i.e. maximum internal detection efficiency) can be realized only below a relatively low temperature, of the order of ~ $0.5 T_c$.[18, 24] Then, the range of saturated bias current $I_{th}$ above which count rates stay constant, increases with further decreasing the temperature. At high temperatures, however, no saturation can be observed. Unlike these optical detectors, our X-ray detector is able to work even at $T = 4.8$ K, just below its critical temperature, with $I_{th}$ as low as 25% of its critical current (see Fig. 3). With the decrease of the temperature, the saturation current $I_{th}$ decreases from ~ 24% of $I_c$ at 4.8 K to ~ 5% of $I_c$ at 1.8 K.

The maximum count rate at $V_A$ = 50 kV and $I_x$ = 1 mA is approximately 3100 s$^{-1}$. Taking the detector area into consideration, the saturated CPSPSM is 2.7 × 10$^6$ cps at all temperatures, even at 4.8 K. The saturated CPSPSM of our Nb and TaN based X-SNSPD at the same acceleration voltage and source current were 2 × 10$^5$ cps at 0.2 $T_c$ and 2.1 × 10$^6$ cps at 0.28 $T_c$ (see Table I), respectively. Therefore the WSi based X-SNSPD indeed has a better detection performance. Apart from the higher photon absorption probability, we believe that the larger hot-spot size in WSi also significantly contributes to the better performance. The hotspot shape (normal state area created by the incident photons) can here be regarded as a cylinder, and we can therefore roughly estimate the hotspot size of our detector using $I_{th}/I_c = (w-D)/w$.[25] According to the extracted threshold current $I_{th}$ from Fig. 3, the calculated hotspot diameters are ~ 700 nm at 4.8 K and ~874 nm at 1.8 K. Note that due to the ultrahigh photon energy, this estimated hotspot diameters here are significantly larger than the determined hotspot diameter induced by the visible photons in two dimensional $W_{0.85}Si_{0.15}$ thin films in ref. 21. For comparison, in Nb and TaN based X-SNSPD, the calculated hotspot size is ~ 420 nm at 1.75 K and ~ 540 nm at 1.85 K, respectively.[12, 13] Thus, the WSi materials are indeed more sensitive to photons with the same energy, and in order to improve the signal to noise ratio, we could in principle even further increase the nanowire width. Taking the X-ray photon flux into consideration, the system detection efficiency of the WSi detector is estimated to be 7.5%.

With respect to the dark counts, our oscilloscope did not collect any detection events within 1 min at the highest temperature and the highest applied bias current (99% of the critical current $I_c$ at $T$ = 4.8 K), and therefore the dark count rates for our detector is lower than ~ 10$^{-2}$ s$^{-1}$. This is not surprising since the chosen device geometry is unfavorable with respect to all the known mechanisms producing dark events.[26] For dark counts induced by the phase-slip centers, either by quantum phase-slips or by thermally activated phase slips, the dark count rate will be significant only when the cross section of the nanowire $wd$ is of the order of $\xi^2$ where $\xi$ = 6.7 nm is the coherence length,[21, 26] which is not the case here. Moreover, the characteristic vortex-antivortex pair unbinding energy is proportional to the film thickness $d$, and therefore the dark counts coming from possible unbinding of vortex-antivortex pairs are intrinsically suppressed for superconducting films with thickness $d \gg \xi$,[27] where we have here. Finally, dark counts stemming from vortices overcoming the vortex-entry energy barrier, are also suppressed since the energy barrier for the vortex entry also scales with the film thickness and is negligible in thick enough films.[26, 27]

The kinetic inductance $L_k = \mu_0 \lambda_{eff} l/w \sim 9.7$ nH where $\lambda_{eff}$ is the effective magnetic penetration depth, $l$ and $w$ denote the total length and width of our detector),[12] leads to a relatively fast recovery ($L_k/R_L \sim 0.2$ ns) of the bias current,[28] where $R_L = 50\ \Omega$ is the load impedance. The detector, however, will latch into a resistive state and is no longer sensitive to the incident photons, as soon as $L_k/R_L$ is comparable with the recovery time of superconductivity. Our detector suffered from the latching problem at all temperatures, and therefore the detection efficiency curves in Fig. 3 terminate at a current which is defined as the latching current $I_L$. Figure 4 shows the measured latching current dependence on the bath temperature under different X-ray photons fluxes. The latching current is independent of the X-ray photon flux, and increases with decreasing temperature since it is determined only by the competition between the phonon escape time and the superconducting current recovery time (determined by the kinetic inductance). As a consequence, the latching effects can, in principle, be significantly reduced by further increasing the kinetic inductance, namely the total length of the nanowire.

In conclusion, we have shown that WSi based superconducting single X-ray photon detectors have a great potential for single X-ray photon detection applications. They can operate with optimum internal detection efficiency just below the critical temperature ≈ 5 K, even above the boiling temperature of liquid helium. The detector reacts to incident photons at ultralow bias currents of less than 1% of the critical current at low temperatures, with a wide saturation range of constant count rate. Dark counts are negligible up to the highest investigated bias currents (99% of $I_c$) and operating temperature (4.8 K). Despite a certain tendency for latching, changes in device design are expected to improve the detector performance even further.

TABLE I. Comparison between $W_{0.8}Si_{0.2}$ and previous Nb and TaN detectors

| Sample | $T_c$ (K) | $w$ (nm) | size (μm²) | CPSPSM (cps) | rise time (ps) | hotspot (nm) | $I_L/I_c$ | $I_{th}/I_c$ |
|---|---|---|---|---|---|---|---|---|
| Nb | 8.40 | 360-410 | 131×55 | 0.2 | 250 | 420 | 5.5% @1.75K | —— |
| TaN-A | 6.70 | 275-340 | 35×33 | 2.1 | 750 | 540 | 52% @1.85K | 8% @1.85K |
| TaN-B | 7.00 | 1800-1900 | 66×119 | 1.3 | 910 | —— | 32% @1.85K | —— |
| $W_{0.8}Si_{0.2}$ | 4.97 | 920 | 41.6×28 | 2.7 | 400 | 874 | 54% @4.8K | 5% @1.8K |

(a)

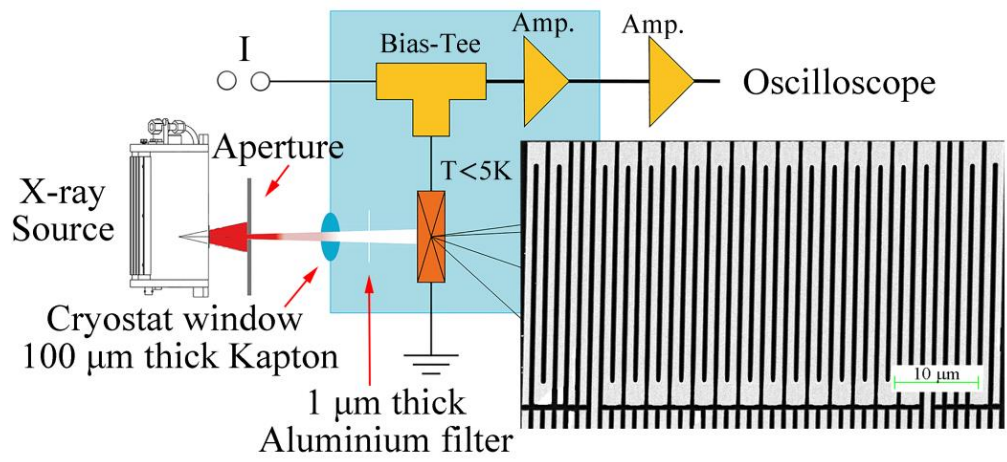

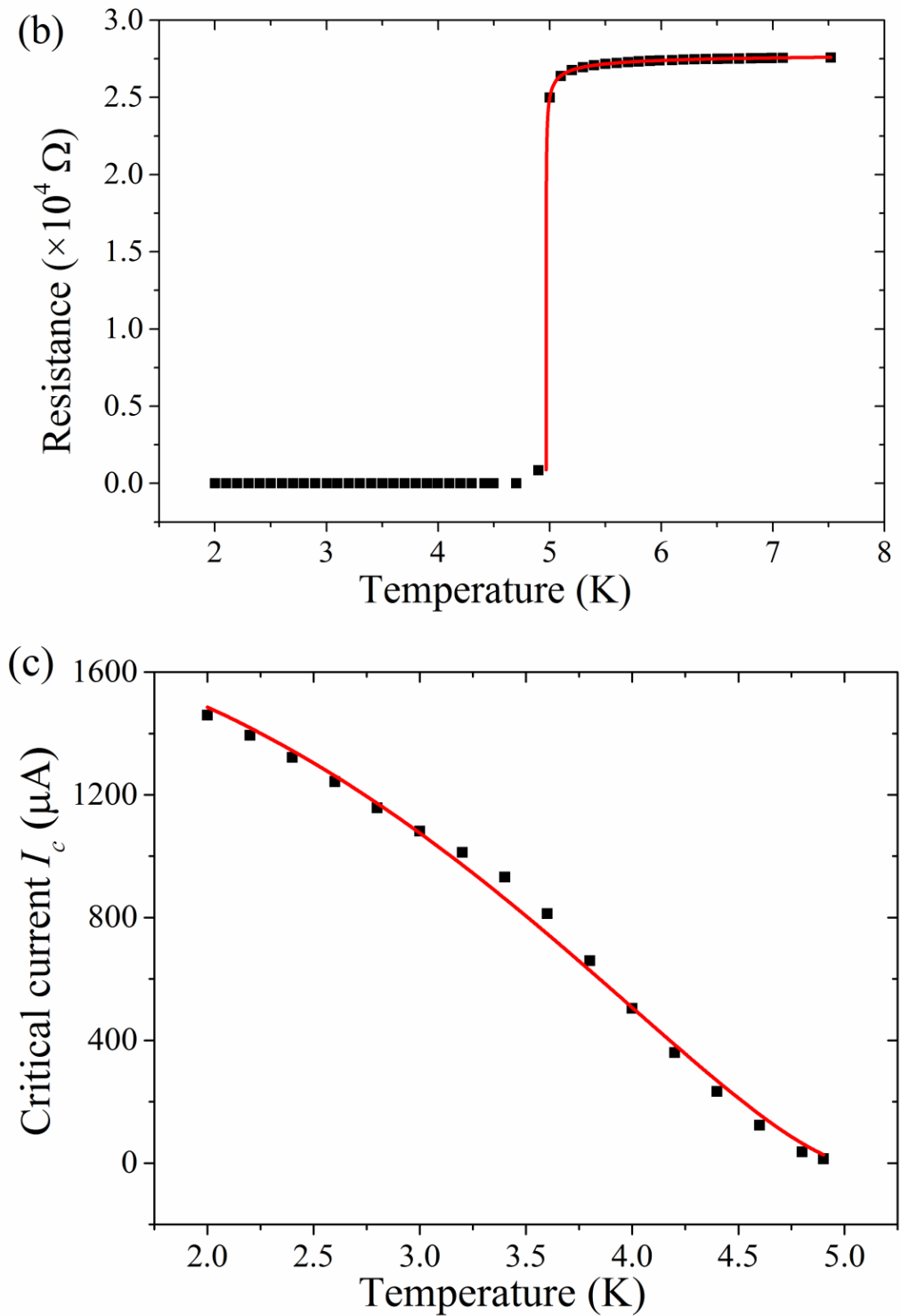

Fig. 1. (a) Schematic of the experimental setup. Inset: SEM image of the X-SNSPD. The total gain of the two amplifiers is 60 dB. (b) Temperature dependence of the measured total resistance of the detector. The red curve is a fit to the Aslamazov-Larkin (AL) fluctuation conductivity formula. (c) The critical current as a function of temperature. The red curve is fitted according to the GL mean field theory.

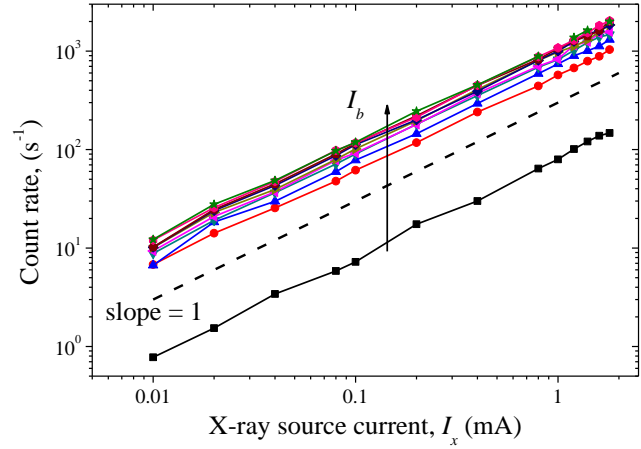

Fig. 2. The measured count rates vs the X-ray source current $I_x$ (with an acceleration voltage $V_A = 30$ kV) at bias current $I_b$ ranging from 10 μA to 100 μA, with an increment of 10 μA. The dash line shows a slope of 1.

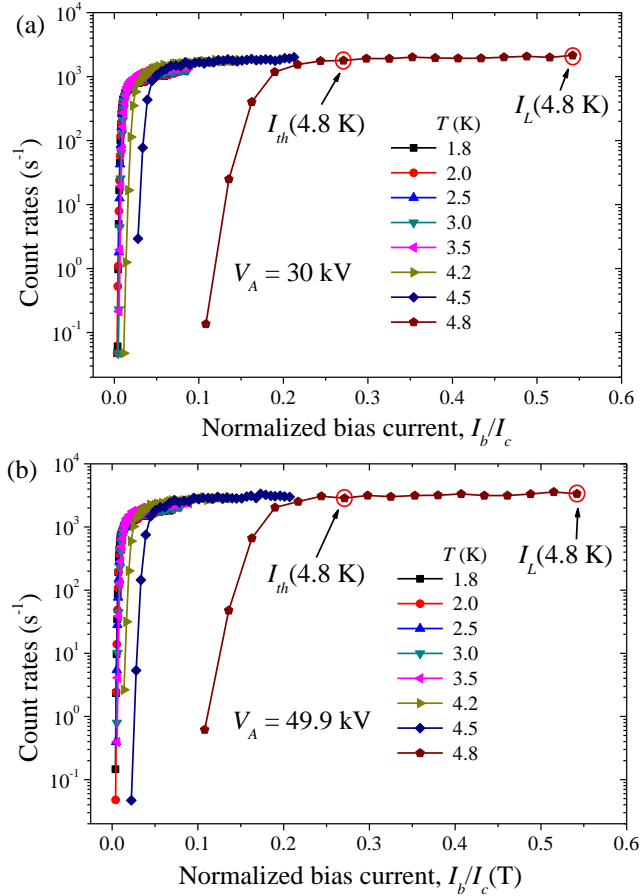

FIG. 3. Count rate as functions of the reduced bias current (normalized by the critical current at each temperature) with acceleration voltage of (a) 30 kV and (b) 49.9 kV. The definitions of the threshold current $I_{th}$ and the latching current $I_L$ at $T = 4.8$ K are circled in the figure.

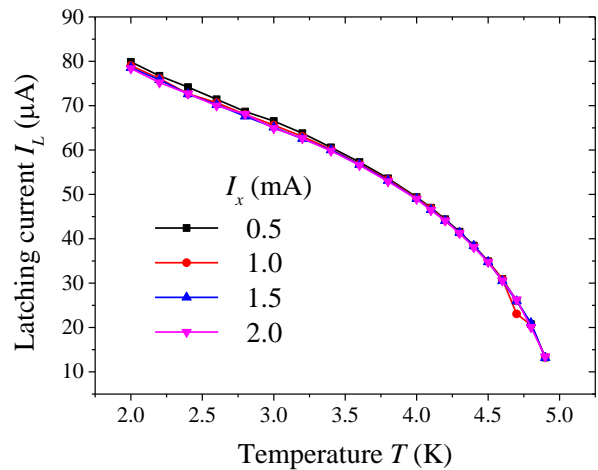

FIG. 4. The latching current $I_L$ as a function of temperature under different X-ray source currents $I_x$.